\documentclass[prl,twocolumn,preprintnumbers,amsmath,amssymb,showpacs]{revtex4}
\usepackage{psfig}

\begin{document}

\title{ Critical behavior and scaling in trapped systems}

\author{Massimo Campostrini and Ettore Vicari} 
\address{Dipartimento di Fisica dell'Universit\`a di Pisa and
I.N.F.N., Largo Bruno Pontecorvo 2, I-56127 Pisa, Italy} 

\date{April 2, 2009}

\begin{abstract}
  We study the scaling properties of critical particle systems confined by a
  potential.  Using renormalization-group arguments, we show that their
  critical behavior can be cast in the form of a trap-size scaling, resembling
  finite-size scaling theory, with a nontrivial {\em trap critical exponent}
  $\theta$, which describes how the correlation length $\xi$ scales with the
  trap size $l$, i.e., $\xi\sim l^\theta$ at $T_c$.  $\theta$ depends on the
  universality class of the transition, the power law of the confining
  potential, and on the way it is coupled to the critical modes.  We present
  numerical results for two-dimensional lattice gas (Ising) models with
  various types of harmonic traps, which support the trap-size scaling
  scenario.
\end{abstract}


\maketitle

% ========================= BODY =========================
%\narrowtext

Recent experimental developments, namely the achievement of Bose-Einstein
condensation in dilute atomic vapors of $^{87}$Rb and 
$^{23}$Na~\cite{CW-02,Ketterle-02}, have attracted great interest in
the study of atomic systems in a trapping potential.  In the
Bose-Einstein condensation scenario, a macroscopic number of trapped
bosonic atoms accumulate in a single quantum state and can be
described by a condensate wave function, which naturally provides the
order parameter of the phase transition. The critical behavior of 
a trapped Bose gas has been recently investigated
experimentally~\cite{DRBOKS-07}, observing an increasing correlation length
compatible with the behavior expected at a continuous transition in the
three-dimensional XY universality class, see, e.g., Ref.~\cite{PV-02} and
references therein, which is also the universality class of the superfluid
transition in $^4$He, see, e.g., Ref.~\cite{Lipa-etal-96}.  However, the
inhomogeneity due to the trapping potential strongly affects the phenomenology
of phase transitions observed in the absence of a trap.  For example,
correlation functions of the critical modes are not expected to develop a
diverging length scale in a trap.  Therefore, a theoretical description of the
critical correlations in systems subject to confining potentials, and of how
they unfold approaching the transition point, is of great importance for
experimental investigations of the critical behavior of systems of trapped
interacting particles.

We consider a trapping potential
\begin{equation}
U(r) = v^p |\vec{r}|^p \equiv (|\vec{r}|/l)^p,
\label{potential}
\end{equation}
where $v$ and $p$ are positive constants and $l = v^{-1}$ is the {\em trap
  size}, coupled to the particle number.  Harmonic potentials, i.e., $p=2$,
are usually realized in experiments.  The effect of the trapping potential is
to effectively vary the local value of the chemical potential, so that the
particles cannot run away.  Let us consider the case in which the system
parameters, such as temperature, pressure and chemical potential, are tuned to
values corresponding to the critical regime of the unconfined system, where
the correlation length diverges as $\xi\sim t^{-\nu}$ and the correlation
function behaves as $G(r)\sim 1/r^{d-2+\eta}$ at $T_c$ ($t\equiv T/T_c-1$ is
the reduced temperature and $\nu$ and $\eta$ are the critical exponents of the
universality class of the transition).  In the presence of a confining
potential, the critical behavior of the unconfined homogeneous system could be
observed around the middle of the trap only in a window where $\xi$ is much
smaller than the trap size but sufficiently large to show the universal
scaling behavior.  If $\xi$ is large but not much smaller than the trap size,
the critical behavior gets somehow distorted by the trap, although it may give
rise to universal effects controlled by the universality class of the phase
transition of the unconfined system, similarly to finite-size scaling in
homogeneous systems of finite size~\cite{FB-72,Barber-83}.  In the present
paper we investigate the critical behavior of trapped systems, putting on a
quantitative ground the above qualitative scenario, and in particular the
relevant effects of the confining potential.  Using renormalization-group (RG)
arguments, we show that the critical behavior of the trapped system can be
cast in the form of a critical {\em trap-size scaling}, resembling standard
finite-size scaling theory for homogeneous systems 
at a continuous transition~\cite{FB-72,Barber-83},
but characterized by a further nontrivial {\em trap critical exponent}.

For the sake of demonstration, as a simple model of trapped particles, one may
consider the $d$-dimensional lattice gas model defined by the Hamiltonian
\begin{equation}
{\cal H}_{\rm Lgas} = 
- 4 J \sum_{\langle ij\rangle}\rho_i \rho_j - \mu \sum_i \rho_i 
+ \sum_i 2 U(r_i) \rho_i
\label{latticegas}
\end{equation}
where the first sum runs over the nearest-neighbor sites of the lattice,
$\rho_i=0,1$ depending if the site is empty or occupied by the particles,
$\mu$ is the chemical potential, and $U(r)$ is the potential
(\ref{potential}).  Far from the origin the potential diverges, thus the
expectation value of the particle number tend to vanish, and therefore the
particles are trapped.  The lattice gas model (\ref{latticegas}) can be
exactly mapped to a standard Ising model, by replacing $s_i=1-2\rho_i$, thus
$s_i=\pm 1$, obtaining
\begin{equation}
{\cal H} = - J \sum_{\langle ij\rangle}s_i s_j - h \sum_i s_i 
- \sum_i U(r_i) s_i
\label{ising}
\end{equation}
where $h=2qJ+\mu/2$ ($q$ is the lattice coordination number).  In the absence
of the trap, this model shows a critical behavior characterized by a diverging
length scale, at the critical point $T=T_c$ and $h=h_c=0$.  Critical
correlations do not develop a diverging length scale in the presence of the
confining potential, i.e., at fixed $v>0$.  We want to study how the critical
behavior is distorted by the trap, and how it is recovered in the limit $v\to
0$.

Our starting point is a scaling Ansatz which extends the scaling law of the RG
theory of critical phenomena~\cite{Wilson-82} (see also
Refs.~\cite{ZJ-book,PV-02}), to allow for the confining potential
(\ref{potential}).  We consider a standard general scenario in which the
transition of the unconfined $d$-dimensional system is characterized by two
relevant parameters, $t\equiv T/T_c-1$ and $h$, as in models
(\ref{latticegas}-\ref{ising}).  We write the scaling law of the singular part
of the free energy density as
\begin{equation}
F(u_t,u_h,u_v,x) = b^{-d} F(u_tb^{y_t},u_hb^{y_h},u_vb^{y_v},x/b)
\label{sfreee}
\end{equation}
where $b$ is any positive number, $x$ is the distance from the middle of the
trap, $u_{t,h,v}$ are scaling fields associated with $t$, $h$, and $v$
respectively.  They are analytic functions of the system parameters, behaving
as $u_t\sim t$, $u_h\sim h$, and $u_v\sim v$ when $t,h,v\to 0$.  $y_{t,h,v}$
are the corresponding RG dimensions: $y_t=1/\nu$ and $y_h= (d+2-\eta)/2$,
while $y_v$ must be determined (see below).  We are neglecting irrelevant
scaling fields, because they do not affect the asymptotic behaviors.  Then,
fixing $u_vb^{y_v}=1$ and introducing the trap size $l=v^{-1}$, we obtain
\begin{equation}
F = l^{-\theta d} {\cal F}(u_tl^{\theta y_t},u_hl^{\theta y_h},xl^{-\theta})
\label{freee}
\end{equation}
where $\theta \equiv 1/y_v$ is the {\em trap exponent}.  From
Eq.~(\ref{freee}) one can derive the trap-size scaling of other observables.
A generic quantity $S$ is expected to behave asymptotically,
when $|t|\to 0$ and in the large-$l$ limit, as
\begin{equation}
S = l^{-\theta y_s} f_s(t l^{\theta/\nu}) = 
l^{-\theta y_s} \bar{f}_s(\xi l^{-\theta}),
\label{genscal}
\end{equation}
where $y_s$ is its RG dimension, $f_s$ and $\bar{f}_s$ are universal functions
(apart from normalizations).  $\xi$ is the correlation length, which behaves
as $\xi\sim l^\theta$ at $T_c$, and as $\xi\sim t^{-\nu}$ when $l\to \infty$.
Note that $\bar{f}_s(x)\sim x^{-y_s}$ for $x\to 0$, so that $S\sim \xi^{-y_s}$
when $l\to\infty$, which is the scaling behavior in the absence of the trap.
These results are quite general, they apply to the lattice gas model
(\ref{latticegas}), as well as to more general interacting gas systems,
fluids, etc...

The trap exponent $\theta$ can be computed by analyzing the RG properties of
the corresponding perturbation at the critical point.  In the case of the
model (\ref{latticegas}), it can be represented by $P_U = \int d^d x\,U(x)
\phi(x)$, where $\phi(x)$ is the order-parameter field of the $\phi^4$ theory
which describes the behavior of the critical modes, see, e.g.,
Ref.~\cite{ZJ-book}.  Since the RG dimensions of the potential $U(x)$ and the
field $\phi$, respectively $y_U=py_v-p$ and $y_\phi=\beta/\nu= (d-2+\eta)/2$,
are related by $y_U + y_\phi = d$, we obtain
\begin{equation}
\theta = {1\over y_v}= {2p \over d+2-\eta+2p}  
\label{rhoh}
\end{equation}
Notice that $\theta\to 1$ when $p\to\infty$, which is consistent with the fact
that when $p\to \infty$ the effect of the trapping potential is equivalent to
confining a homogeneous system in a box of size $L=l$ with fixed boundary
conditions, thus leading to a standard finite size scaling where the RG
dimension of the size $L$ is formally minus one~\cite{FB-72,Barber-83}.

As examples of observables in model (\ref{ising}), we consider the local
magnetization and the energy density in the middle of the trap, i.e.,
\begin{equation}
M_0\equiv \langle s_0 \rangle,\qquad 
E_0=\langle s_0 s_1\rangle
\label{mag0}
\end{equation}
(where 1 is a nearest neighbor of the center site 0),
and the correlation function
\begin{equation}
G_0(r) \equiv \langle s_0 s_r \rangle - 
\langle s_0 \rangle \langle s_r \rangle 
\label{corr}
\end{equation}
In the lattice gas model $M_0$ and $G_0(r)$ are related to the particle
density and its correlations.  Their asymptotic trap-size scaling behaviors
are given by
\begin{eqnarray}
&&M_0 = l^{-\theta \beta/\nu} f_m(t l^{\theta/\nu}) =  
l^{-\theta \beta/\nu} \bar{f}_m(\xi l^{-\theta}),
\label{moscal} \\
&&G_0(r) = l^{-\theta\eta} f_g(t l^{\theta/\nu},rl^{-\theta}) = 
l^{-\theta\eta} \bar{f}_g(\xi l^{-\theta},r/\xi),
\quad \label{gscal}\\
&&E_0 = E_{\rm ns}(t) + l^{-\theta(d-1/\nu)} f_e(t l^{\theta/\nu}),
\label{enescal0}
\end{eqnarray}
where $\beta/\nu=(d-2+\eta)/2$ and $E_{\rm ns}(t)$ is a nonsingular function.
The scaling behavior at $T_c$ can be obtained from the above equations by
setting $t=0$,
\begin{eqnarray}
&&M_0 \sim l^{-\theta\beta/\nu}, 
\quad E_0 - E_c\sim l^{-\theta(d-1/\nu)},
\label{moenescaltc}\\
&&G_0(r) \sim l^{-\theta\eta} g(rl^{-\theta}).
\label{magrscalisi} 
\end{eqnarray}

In order to check the trap-size scaling scenario, we consider the
square-lattice Ising model (\ref{ising}) with $J=1$ (for which we know the
critical temperature and energy, i.e., $T_c=2/\ln(\sqrt{2}+1)$ and
$E_c=\sqrt{2}$, and the critical exponents, $\nu=1$, $\eta=1/4$) in a harmonic
trap, i.e., the potential (\ref{potential}) with $p=2$.  According to
Eq.~(\ref{rhoh}), the corresponding trap exponent is $\theta=16/31$.  Beside
two-dimensional (2D) traps, we also consider one-dimensional (1D) traps, where
the confining potential acts only along one direction, i.e., $U(r) = v^p
|x|^p$ depends only on $x$, while there is translation invariance along the
other $y$ direction.  Note that the trap-size scaling formulae, including the
value of the trap exponent $\theta$, apply to both 2D and 1D traps.

\begin{figure}[tb]
\centerline{\psfig{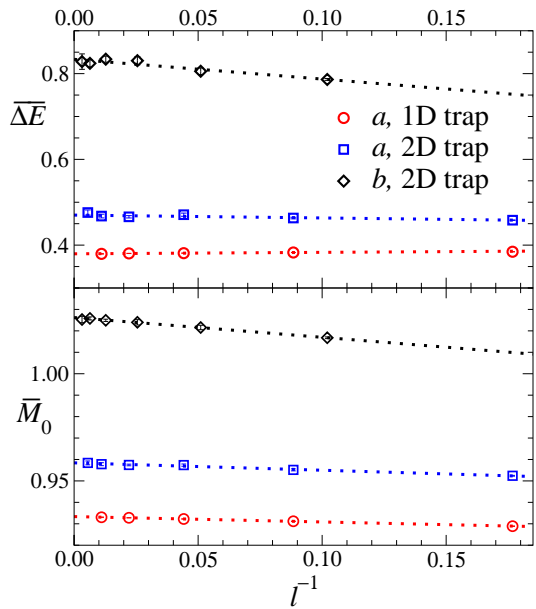}}
\caption{ Trap-size scaling of $M_0$ and $E_0$ at $T_c$: we plot
  $\overline{M}_0 \equiv l^{\theta\eta/2} M_0$ (below) and $\overline{\Delta
    E} \equiv l^{\theta}(E_0-E_c)$ (above) vs $l^{-1}$, for the model
  (\ref{latticegas}) with 1D and 2D harmonic traps (denoted by ``$a$'') for
  which $\theta=16/31$, and the model (\ref{isinge}) with a 2D harmonic trap
  (by ``$b$'') for which $\theta=2/3$. The dotted lines show linear fits.}
\label{magnenescal}
\end{figure}

\begin{figure}[tb]
\centerline{\psfig{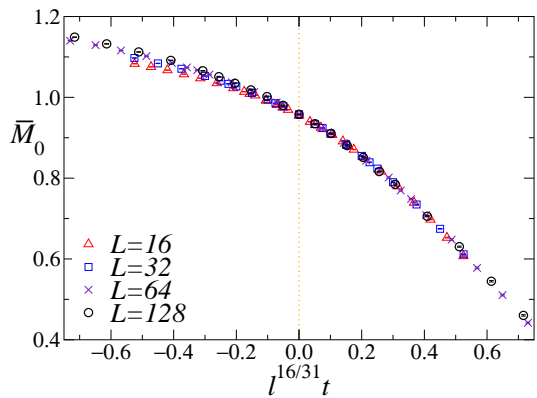}}
% \vspace{2mm}
\caption{
  Trap-size scaling of $M_0$ for the model (\ref{latticegas}) in a 2D trap.
  $\overline{M}_0 \equiv l^{2/31} M_0$ and $L=\sqrt{2}l$.  }
\label{maxtss}
\end{figure}

\begin{figure}[tb]
\centerline{\psfig{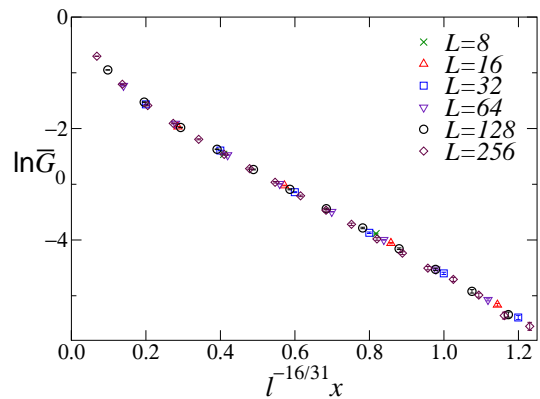}}
% \vspace{2mm}
\caption{
  Scaling at $T_c$ of $G_0(r)$ for the model (\ref{latticegas}) in a 2D trap.
  $\overline{G}_0(x)\equiv l^{4/31} G_0(x)$ and $L=\sqrt{2}l$. }
\label{corrltch}
\end{figure}

We performed several Monte Carlo (MC) simulations around the critical point
and for various values of the trap size $l$.  The lattice size $L$ and
harmonic potential of the simulated systems were chosen to have the spin
variables effectively frozen at the boundary, making unnecessary the use of
larger lattices~\cite{MCdetails}. The MC results for $M_0$ and $E_0$ at $T_c$
provide accurate checks of the trap-size scaling predicted by
Eqs.~(\ref{moenescaltc}).  In Figs.~\ref{magnenescal} we plot
$\overline{M}_0\equiv l^{2/31} M_0$ and $\overline{\Delta E}\equiv l^\theta
(E_0-E_c)$ versus $l^{-1}$, which is the order of the expected leading
corrections to the asymptotic scaling~\cite{scalcorr}.  The lines represent
fits of all available data, i.e., for $L=2^n$ with $n=3,...,8$, to
$\overline{M}_0= a + b/l$ (in all cases with $\chi^2/{\rm dof}\lesssim
1$).~\cite{fit0} In Fig.~\ref{maxtss} we check the scaling of $M_0$ around
$T_c$ in the case of the 2D trap, by plotting $\overline{M}_0\equiv l^{2/31}
M_0$ versus $t l^{16/31}$.  As predicted by Eq.~(\ref{moscal}), the data
approach a scaling function $f_m(tl^{16/31})$ in the large-$l$ limit, where
scaling corrections are suppressed.  Fig.~\ref{corrltch} shows results for the
correlation function (\ref{corr}) in the case of the 2D trap, at $T_c$.  They
clearly support Eq.~(\ref{magrscalisi}), indeed the data of
$\overline{G}_0(r)\equiv l^{4/31}G_0(r)$ follow the same scaling function
$g(rl^{-16/31})$ for all trap sizes (scaling corrections are very small and
not visible in Fig.~\ref{corrltch}).  $G_0(r)$ appears to decay rapidly
at large distance, with a length scale 
behaving as $\xi \sim l^\theta$.

\begin{figure}[tb]
\centerline{\psfig{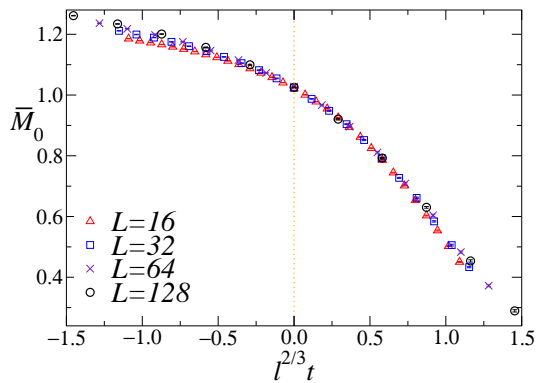}}
% \vspace{2mm}
\caption{
Trap-size scaling of $M_0$ for the model (\ref{isinge}) in a 2D trap.
$\overline{M}_0 \equiv l^{1/12} M_0$  and $L=\sqrt{2/3}l$.
}
\label{mattss}
\end{figure}

\begin{figure}[tb]
\centerline{\psfig{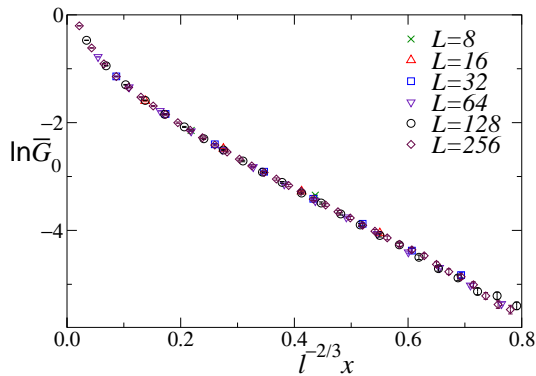}}
% \vspace{2mm}
\caption{ 
  Scaling at $T_c$ of $G_0(r)$ for the model (\ref{isinge}).  
  $\overline{G}_0(x)\equiv l^{1/6} G_0(x)$ and $L=\sqrt{2/3}l$.}
\label{corrltcht}
\end{figure}

To further check the trap-size scaling scenario, we also consider another
Ising-like model with the confining potential (\ref{potential}) coupled to the
energy density, i.e.,
\begin{equation}
{\cal H}_e = - J \sum_{\langle ij\rangle} [1+U(r_{ij})]  
s_i s_j - h \sum_i s_i 
\label{isinge}
\end{equation}
where the first sum runs over nearest-neighbor sites, and $r_{ij}\equiv
(r_i+r_j)/2$. Also in this case the spin variables get frozen at large
distances. More precisely, ${\rm lim}_{h\to 0^+} {\rm
  lim}_{|r|\to \infty}\langle s_r \rangle = 1$ at any $T$, corresponding to
vanishing particle density in the lattice gas model.  The effects of the
confining potential at the transition can be again described by a trap-size
scaling, cf. Eqs.~(\ref{moscal}-\ref{magrscalisi}), but with a different trap
exponent $\theta$.  Indeed, a RG analysis of the perturbation arising from the
potential $U$ in model (\ref{isinge}), $Q_U = \int d^dx\,
U(x) \phi^2(x)$, gives~\cite{phi2p} 
\begin{equation}
\theta = {p \nu\over 1+p\nu}  
\label{rhoe}
\end{equation}
We again consider a square-lattice model with a harmonic 2D trap.  Since
$\nu=1$, Eq.~(\ref{rhoe}) gives $\theta=2/3$, which differs from
the value 16/31 for the model (\ref{ising}).  We performed MC simulations
similarly to the previous case. Results for the trap-size scaling of
the magnetization $M_0$ and the energy $E_0$ at the origin, and the
correlation function (\ref{corr}) are shown in Figs.~\ref{magnenescal},
\ref{mattss} and \ref{corrltcht}.  They again fully support the predicted
trap-size scaling behaviors.~\cite{fitisie} 

In conclusion, we have shown that the critical behavior of systems in a
confining potential can be described by a universal trap-size scaling
(expected to be largely independent of the microscopic details of the model),
characterized by a trap exponent $\theta$ which describes how the correlation
length scales with the trap size, $\xi\sim l^\theta$ at $T_c$.  The exponent
$\theta$ essentially depends on the universality class of the transition, the
power law of the confining potential, and on the way it is coupled to the
critical modes.  These results are very general, and by no means limited to
two dimensions. 

We finally discuss the critical behavior of a 3$d$ interacting Bose gas
trapped by a harmonic potential, which has been recently investigated
experimentally~\cite{DRBOKS-07}.  This system is expected to undergo a
continuous transition in the 3$d$ XY universality class, characterized by a
complex order parameter $\phi(x)$ with $U(1)$ symmetry.  The confining
potential $U(x)$ is coupled to the particle density.  The corresponding
perturbation is~\cite{DSMS-96} $Q_U=\int d^3 x\, (|x|/l)^2 |\phi(x)|^2$, where
$\phi(x)$ is the order-parameter field.  The same RG arguments which led us to
Eq.~(\ref{rhoe}) give $\theta = 2 \nu/ (1+2\nu)$, thus $\theta=0.57327(4)$
using~\cite{CHPV-06} $\nu=0.6717(1)$. This implies that one can neglect the
trap effects only when the correlation length satisfies $\xi\ll
l^{\theta}=l^{0.573}$.  The experimental results of Ref.~\cite{DRBOKS-07} were
likely taken in this region, and led to the estimate $\nu=0.67(13)$, by
fitting the data to the standard behavior $\xi\sim t^{-\nu}$.  For larger
values of $\xi$, when $\xi$ and $l^\theta$ become comparable, the trap-size
scaling discussed in this paper is expected to provide the correct critical
behavior.  We believe that experiments probing the trap-scaling regime can be
very interesting, analogously to experiments probing finite-size scaling
behavior in $^4$He at the superfluid transition~\cite{GKMD-08}.  Moreover,
accurate studies of the critical properties of trapped systems, to check
universality and determine the critical exponents, require a robust control
of the effects of the confining potential. In this respect, one may actually
exploit trap-size scaling, using it to infer the critical exponents from the
data, analogously to finite-size scaling techniques for the accurate
determination of the critical parameters, see, e.g., Ref.~\cite{PV-02}.

Helpful discussions with E. Arimondo, P. Calabrese and M. Oberthaler are
gratefully acknowledged.


\begin{thebibliography}{99}
\bibitem{MCdetails} In the case of the 2D trap we considered lattices with $-L
  \le x,y \le L$, and trap potential $U_L(x) = c (x^2+y^2)/L^2$, while at the
  boundaries $|x|,|y|=L+1$ the spin variable was kept fixed: $s_i=1$.
  In most simulations we fixed $c=2$, i.e., $L=\sqrt{2}l$.  In the
  case of the 1D trap, we considered 
  $L_x\times L_y$ lattices with $U_L(x) = c x^2/L_x^2$, and $L_y/L_x\gg 1$
  with periodic boundary conditions along $y$.
\bibitem{scalcorr} Leading scaling corrections arise from nonlinear dependence
  of the scaling field $u_v$ on $v$, i.e., its $O(v^2)$ term.
  Corrections due to irrelevant perturbations are more suppressed,
  because their leading RG exponent is $\omega=2$ in the two-dimensional Ising
  model~\cite{PV-02}.
\bibitem{phi2p} $\phi^2(x)$ represents the energy density operator in the
  corresponding $\phi^4$ theory, i.e. $H = \int d^d x
  \left[ (\partial_\mu \phi)^2 + r \phi^2 + u \phi^4\right]$. We obtain
  Eq.~(\ref{rhoe}) from the RG relation $y_U + y_{\phi^2} = d$ and the fact
  that the RG dimension of $\phi^2(x)$ is $y_{\phi^2}=d-y_t$.
\end{thebibliography}
\end{document}